%% file: pion.tex
\documentclass[12pt]{article}
\usepackage{amsmath,epsfig,amssymb,eufrak,cite}

\date{}

\newcommand{\e}{{\rm{e}}}
\newcommand{\msb}{\overline{\mathrm{MS}}}
\newcommand{\dds}{\stackrel{\leftrightarrow}{D}}

\input macros.sty      
\input eqalign.sty     

\begin{document}

\begin{titlepage}

\title{
  {\vspace{-0cm} \normalsize
  \hfill \parbox{40mm}{DESY/04-087\\
                       SFB/CPP-04-14\\May 2004}}\\[30mm]
Non-perturbative Pion Matrix Element of a
twist-2 operator from the Lattice} 
  \author{M.\ Guagnelli$^{a}$, K.\ Jansen$^{b}$, F.\ Palombi$^{a,c}$,\\ 
   R.\ Petronzio$^{a}$, A.\ Shindler$^{b}$ and I.\ Wetzorke$^{b}$\\
\\
   {Zeuthen-Rome (ZeRo) Collaboration}\\
\\ 
  {\small $^a$ Dipartimento di Fisica, Universit\`a di Roma 
        {\em Tor Vergata}}\\ 
  {\small and INFN, Sezione di Roma II,} \\
  {\small Via della Ricerca Scientifica 1, I-00133 Rome, Italy} \\
  {\small $^b$ NIC/DESY Zeuthen, Platanenallee 6, D-15738 Zeuthen,
   Germany}\\
  {\small $^c$ {\em E.~Fermi} Research Center,
   c/o Compendio Viminale, pal.~F, I-00184 Rome, Italy} \\
}

\maketitle

\begin{abstract}
We give a continuum limit value of the lowest moment of a twist-2
operator in pion states from non-perturbative lattice calculations.
We find that the non-perturbatively obtained 
renormalization group invariant matrix element 
is $\langle x\rangle_{\mathrm{RGI}} = 0.179(11)$, which corresponds
to $\langle x\rangle^{\msb}(2\mathrm{~GeV}) = 0.246(15)$.  
In obtaining the renormalization group invariant matrix element, 
we have controlled important systematic errors
that appear in typical lattice simulations, such as non-perturbative
renormalization, finite size effects and effects of a 
non-vanishing lattice spacing. The crucial limitation of our
calculation is the use of the quenched approximation. 
Another question that remains not fully clarified is the chiral
extrapolation of the numerical data. 
\vspace{0.8 cm}
\noindent
\end{abstract}

\end{titlepage}

\section{Introduction}

Deep inelastic scattering \cite{thebook} 
continues to provide important information
on the structure of hadrons. Phenomenological fits to 
experimental data give values for the moments of parton 
distribution functions (PDF), including estimates of their errors, see
e.g. \cite{alekhin,martin,blume}.
Since such moments can be expressed as expectation values of 
local operators, they are accessible to lattice calculations
\cite{stefanoreview}.
A direct comparison of these lattice calculations of moments 
with the results of the phenomenological fits will test whether
these fits are consistent with direct QCD predictions. 
If we think of, e.g.,  precise determinations of the strong coupling constant
from scaling violations in 
deep inelastic scattering, such non-trivial checks are
mandatory.

Lattice results do not come for free, however: 
Concepts of non-perturbative renormalization have to be developed;
in the process of 
moving from the continuum of space time to an euclidean lattice
a non-vanishing value of a lattice spacing $a$ is introduced, leading 
to discretization effects;
running simulations on a computer necessitates the use of an
only finite volume; 
the limited amount of computing resources leads 
to the fact that simulations are performed at rather large
values of the quark masses that are far from their values assumed in
nature. 
In addition, at present we are left with the quenched 
approximation, neglecting internal quark degrees of freedom.
Finally, the numerical results are plagued by statistical errors
that can be substantial for bad choices of operators.

In the course of our work 
\cite{ref:perturbative,ref:non-pert,ref:universal,ref:invariant,ref:matrixele,rgissf} 
to reach a reliable value for a moment
of a parton distribution function we eliminated important 
sources of systematic errors besides the quenched approximation
(see \cite{karl_osaka,andrea_elba,karl_genua} for summaries of these
works). 
The transition to full dynamical simulations is undertaken world-wide
today \cite{karlreview}
and the next years will see the exciting results of such calculations.
Another open question is the chiral extrapolation that is not 
understood presently (see \cite{panel2001} and refs. therein).

Let us sketch how we have addressed the systematic errors of the 
lattice calculations in our work: 

\begin{itemize}
\item {\em Non-perturbative renormalization}\\
We adopt in our work \cite{ref:perturbative,ref:non-pert,rgissf} 
the Schr\"odinger functional (SF) scheme \cite{sfscheme}, 
which has been
proved to be a very successful and practical way to compute
scale dependent renormalization constants. By evolving deep into
the perturbative regime, renormalization group invariant (RGI) quantities
can be determined that allow to relate lattice results into any
desired continuum renormalization scheme 
\cite{rainer_review,martin_review,rainertalk}.
In this paper we compute the renormalization factor
at the matching scale (see sect. \ref{sec.:3.2} for details).
\item {\em Discretization effects}\\
We have controlled effects of a non-vanishing lattice spacing 
$a$ by performing all our lattice calculations with several values of $a$
with two different
lattice formulations of lattice QCD: one is the standard 
Wilson fermion, the other the non-perturbatively improved Wilson
fermion formulation. In this way, all quantities could be 
extrapolated to their continuum value in a controlled way 
performing a constrained continuum limit \cite{ref:universal,{rgissf}}.
In particular in this work we have used this strategy to 
control the continuum limit of the renormalized matrix element.
\item {\em Finite Size Effects}\\
The finite size effects were controlled in two ways. For the 
computation of the evolution of the renormalization constants
\cite{ref:non-pert,ref:invariant,rgissf},
the finite volume Schr\"odinger functional scheme was used, in
which the scale $\mu$ is identified with the inverse 
lattice extent $L^{-1}$. In this way, the finiteness of the lattice
has been utilized to determine the scale dependence. 

For the matrix elements, a careful test of finite size 
effects has been performed \cite{fse} with the somewhat surprising result
that such effects are rather large for the pion matrix element at a point
where the pion mass itself shows no effects. Since
this was done for pion matrix elements only, it will be very 
important to repeat the analysis for baryon matrix elements, where
such effects are expected to be even larger \cite{baryonfse}. 
\item {\em Statistical errors}\\
By employing generalized boundary conditions in space, 
the signal to noise ratio of operators that need an external
momentum could be minimized by optimizing parameters that 
are introduced by the generalized boundary conditions \cite{rgissf}. 
This allowed us to obtain results that would not have 
been possible otherwise. 
\item {\em Chiral extrapolation}\\
The extrapolation of the numerical data from the rather heavy 
quark masses, where the simulations are performed, to their 
physical values is still under debate and not clarified,
see \cite{panel2001} for a general discussion,
and \cite{{thomas1},{australians},{savage1},{savage2},{savage3},{chen1}} for works
on the chiral extrapolation of PDF.
We will follow the strategy in this work that we first extrapolate
the values of our non-perturbatively improved matrix elements
to the continuum limit. Comparisons to predictions of chiral
perturbation theory are then performed directly in the continuum
in order to avoid lattice spacing effects that would render
the interpretation difficult.
\end{itemize}

After having understood and overcome the above difficulties, we can now
provide results that are free of these systematic uncertainties 
besides the, presently still unavoidable, quenched approximation.  
In this work we summarize our result and, most importantly, we provide
the missing part, i.e. the renormalized matrix element at the hadronic scale 
$\mu_0\simeq 275~ {\rm MeV}$ down to which the scale evolution has been 
computed (\cite{rgissf}), with a full control on the continuum limit
and a carefull analysis of the chiral extrapolation.
As the most important quantity we consider the {\em renormalization group invariant}
(RGI) matrix element of the twist-2 operator between pion states corresponding to the
first moment of the valence quark distribution, which we find

\be
\langle x\rangle_\mathrm{RGI} = 0.179(11) .
\label{rgiresult}
\ee

This quantity is of central importance since it allows to relate the 
non-perturbati\-vely obtained results using a particular lattice 
renormalization scheme to more 
conventional schemes. For example, if we use the $\msb$ scheme, we find that 
at a scale of 2 GeV the value of the matrix element compares to
phenomenological estimates of the same quantities 
extracted from global fits of experimental
data \cite{experiment} as follows 
\footnote {Some details on the extraction of the 
phenomenological number can be found in sect. \ref{sec.:5.1} of this paper.}:

\begin{eqnarray}
\langle x\rangle^{\msb}(\mu=2\mathrm{~GeV}) & = & 0.246(15) \nonumber \\
\langle x\rangle_{phen}(\mu=2\mathrm{~GeV}) & = & 0.21(2) .
\label{finresults}
\end{eqnarray}

This results demonstrates that the running of moments of parton distribution
functions in the {\em continuum} can be calculated from
lattice simulations in an intrinsic perturbative scheme like the $\msb$
retaining all the non-perturbative information coming from the RGI matrix element. 
In addition, the errors
coming from the lattice simulations are comparable to the experimental
errors, which opens the possibility to perform direct tests of QCD as
the theory of the strong interaction. 

\section{Renormalized matrix element}

The moments of parton density distributions
are related to expectation values of local
operators, which are renormalized multiplicatively by applying
appropriate
renormalization factors $Z(\mu)$ that depend
on the energy scale $\mu$ (see, e.g. ref.~\cite{thebook}).
This leads to consider renormalized
matrix elements $O^{\mathrm{ren}}(\mu)$ using some 
renormalization scheme, denoted by {\it ren}. 
For lattice calculations, we are aiming at
here, a very useful scheme is 
the Schr\"odinger function (SF) renormalization scheme \cite{sfscheme} since
it applies in small volumes. 

If the energy scale $\mu$ of $Z(\mu)$ 
is chosen large enough, it is to be expected, and indeed
it can be checked explicitly,
that the scale evolution is very well described by perturbation theory,
giving rise to the following definition  of
a {\em renormalization group invariant} (RGI) matrix element:
\begin{equation}
O_{\mathrm{RGI}} = O^{\mathrm{SF}}(\mu)\cdot
f^{\mathrm{SF}}(\bar{g}^2(\mu))
\label{eq:inv}
\end{equation}
with $\bar{g}(\mu)$ the running coupling computed in the same SF scheme and
\begin{equation}
  f^{\mathrm{SF}}(\bar{g}^2) = (\bar{g}^2(\mu))^{-\gamma_0/2b_0}
  \cdot \exp\left\{
 -\int_0^{\bar{g}(\mu)} dg\left[\frac{\gamma(g)}{\beta(g)} 
- \frac{\gamma_0}{b_0g}
\right]\right\} ,
\label{eq:f}
\end{equation}
where $\beta(g)$ and $\gamma(g)$ are the $\beta$ and 
anomalous-dimension functions
computed to a given order of perturbation theory in a specified scheme, i.e.
here the SF scheme. Once we know the value of 
$O_{\mathrm{RGI}}$ evaluated
non-perturbatively, the running matrix element in a preferred 
scheme can be computed,
for example in the $\msb$ scheme:
\begin{equation}
O^{\mathrm{\overline{\mathrm MS}}}(\mu) =
O_{\mathrm{RGI}}/f^{\overline{\mathrm{MS}}}({g}_{\overline{\mathrm{MS}}}^2(\mu))
\end{equation}
with now, of course, the $\beta$ and $\gamma$ functions computed in the
$\msb$ scheme. Thus, although the SF is an unphysical finite volume scheme
(but therefore most suited for lattice simulations) it can be related to more
conventional continuum renormalization schemes by providing 
the renormalization group invariant quantities.  

A non-perturbatively obtained value of the
renormalization group invariant matrix element is hence of central
importance.
Its calculation
has to be performed in several steps. The reason is that we have to cover
a broad range of energy scales -- from the deep perturbative to the
non-perturbative region.
Using the scale dependent renormalization factor $Z^{\mathrm{SF}}(\mu)$, we        
write the renormalized matrix element
of eq.~(\ref{eq:inv}) as
\begin{equation}
O^{\mathrm{SF}}(\mu) = 
\frac{\langle h | \mathcal{O}|h\rangle}
         {Z^{\mathrm{SF}}( \mu )} ,
\label{eq:split}
\end{equation}
where $|h\rangle$ is the hadron state we are interested in. 
So far, all our discussions have been in the continuum. However, if we think
of the lattice regularization and eventual numerical simulations to obtain
non-perturbative results, it would be convenient to compute the 
renormalized matrix
element at only one (i.e. small hadronic) scale $\mu_0$.
We therefore rewrite the r.h.s. of eq.~(\ref{eq:split}) as
\begin{equation}
           \frac{\langle h | \mathcal{O}|h\rangle}
           {Z^{\mathrm{SF}}( \mu )} = \frac{\langle h | \mathcal{O}| h \rangle}
           {Z^{\mathrm{SF}}( \mu_0 )}
           \cdot \underbrace{\frac{Z^{\mathrm{SF}}( \mu_0 )}
           {Z^{\mathrm{SF}}( \mu )}}_{\equiv
           \sigma( \mu/\mu_0, \bar{g}(\mu)) } ,
\end{equation}
where we
introduce the {\rm step scaling function} $\sigma( \mu/\mu_0, \bar{g}(\mu))$,
which describes the
evolution of the renormalization factor from a scale $\mu_0$ to a scale $\mu$.
The advantage of concentrating on the step scaling function instead of the
renormalization factor itself is that the step scaling function is well defined
in the continuum and hence suitable for eventual continuum extrapolations
of lattice results.
We finally write the r.h.s. of eq.~(\ref{eq:inv}) as
\begin{equation}
            O_{\mathrm{RGI}} = O^{\mathrm{SF}}( \mu_0)
           \underbrace{\sigma( \mu/\mu_0, \bar{g}(\mu))
           \cdot f^{ \mathrm{SF} }(\bar{g}^2( \mu ))}_{
           \equiv \EuFrak{S}_{\rm INV}^{\rm UV,SF}(\mu_0)}
\label{pieces_of_oren}
\end{equation}
with $O^{\mathrm{SF}}( \mu_0)$ the renormalized matrix element,
which is to be computed only once at
a scale $\mu_0$ and the ultraviolet (UV) invariant step scaling function
$\EuFrak{S}_{\rm INV}^{\rm UV,SF}(\mu_0)$, 
which still depends on the infrared scale $\mu_0$, and on the renormalization scheme
adopted.
In ref.~\cite{rgissf} we have given a value for the 
UV invariant step scaling function and checked the independence on the ultraviolet
scale $\mu$. 
In this work we will provide the
missing part, i.e. the renormalized matrix element at the scale
$\mu_0$. 
The final result for the renormalized matrix element of the pion
that we are going to provide here, was made possible by a number of theoretical
and conceptual developments that we could achieve over the last years
\cite{ref:perturbative,ref:non-pert,ref:universal,ref:invariant}. 
The methods developed there can immediately
be taken over to other matrix elements than the lowest twist 
case of the pion considered here and to the unquenched situation. 
Nevertheless, with this paper we want to finish 
the analysis of the pion matrix element with the aim to have eliminated
important systematic uncertainties besides the quenched approximation.

\subsection{Transfer matrix decomposition}

The moments of parton distribution functions (PDF) 
are related to matrix elements of leading twist 
$\tau$ ($\tau=$ dim - spin) operators of given spin, between hadron states $h(p)$
\bea
\hspace*{-5mm}
\langle h(p)|{\cal O}_{\mu_1 \ldots \mu_N}|h(p) \rangle &=& M^{(N-1)}(\mu)
p_{\mu_1} \cdots p_{\mu_n} \nn \\ 
&& + {\rm{terms~}} \delta_{\mu_i \mu_j}, 
\eea
\be
\hspace*{-7mm}
\langle x^{(N-1)} \rangle (\mu) = M^{(N-1)}(\mu = Q). 
\ee
We concentrate in this
work on the twist-2 operator corresponding to the second moment of the
parton distribution functions (PDF) between charged pion states.
In the following we consider the fermionic fields $\psi(x)$ and $\bar\psi(x)$
as doublets in the flavour space. In particular we will concentrate on the
valence $u$ or $\bar d$ distribution as explained below.
This amounts to consider operators of the form 
\be
{\mathcal O}_{\mu \nu}(x) = \frac{1}{4}\bar\psi (x) \gamma_{\{\mu}
\lrD_{\nu\}} \psi (x) - \delta_{\mu \nu} \cdot {\rm trace~terms}\; ,
\label{eq:ope1_tr}
\ee
where $\{\cdots\}$ means symmetrization on the Lorentz indices,
$\lrD_\mu = \rD_\mu - \lD_\mu$ and
\be
  \rD_\mu = \frac{1}{2}(\nabla_\mu + \nabla^*_\mu), 
  \qquad 
  \lD_\mu = \frac{1}{2}
  (\overleftarrow\nabla_\mu + \overleftarrow\nabla^*_\mu)\; .
\ee
The definitions of the lattice derivatives and conventions are given in the appendix.

There are
two representations of such an operator on the lattice
\cite{h4_first}. The first representation takes $\mu\ne\nu$, 
whereas the second uses $\mu=\nu$. The precise definitions of the
operators used here are 
\begin{equation}
{\cal O}_{12}(x) = \frac{1}{4}
       \bar\psi(x) \gamma_{\{1} \dds_{2\}} \psi(x)
\label{O12}
\end{equation}
and
\be
\cO_{44}(x) = \frac{1}{2} \bar\psi(x) \Big[ \gamma_4 \lrD_4 - \frac{1}{3}
\sum_{k=1}^3 \gamma_k \lrD_k \Big]  \psi(x)\; .
\label{O44}
\ee
In computing the matrix elements of these operators,
a non-zero momentum in two different spatial directions 
has to be supplied for ${\cal O}_{12}(x)$ in
eq.~(\ref{O12}), whereas for the operator $\cO_{44}(x)$ in eq.~(\ref{O44}) 
no momentum is needed. It is to be expected, and indeed verified in numerical 
simulations, that the signal of the matrix element of
$\cO_{44}(x)$ is better than the one of ${\cal O}_{12}(x)$. Thus in the
following investigation we consider only $\cO_{44}(x)$.

\begin{figure}
\vspace{-0.0cm}
\begin{center}
\psfig{file=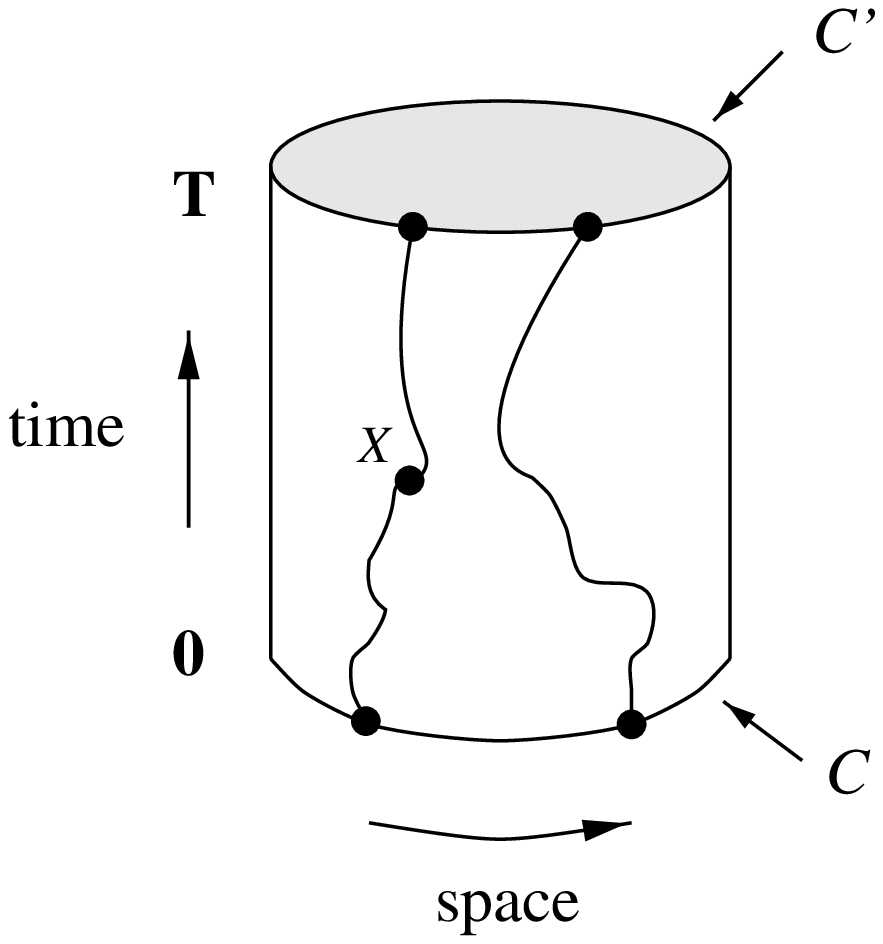,width=5.5cm,height=5.0cm}
\psfig{file=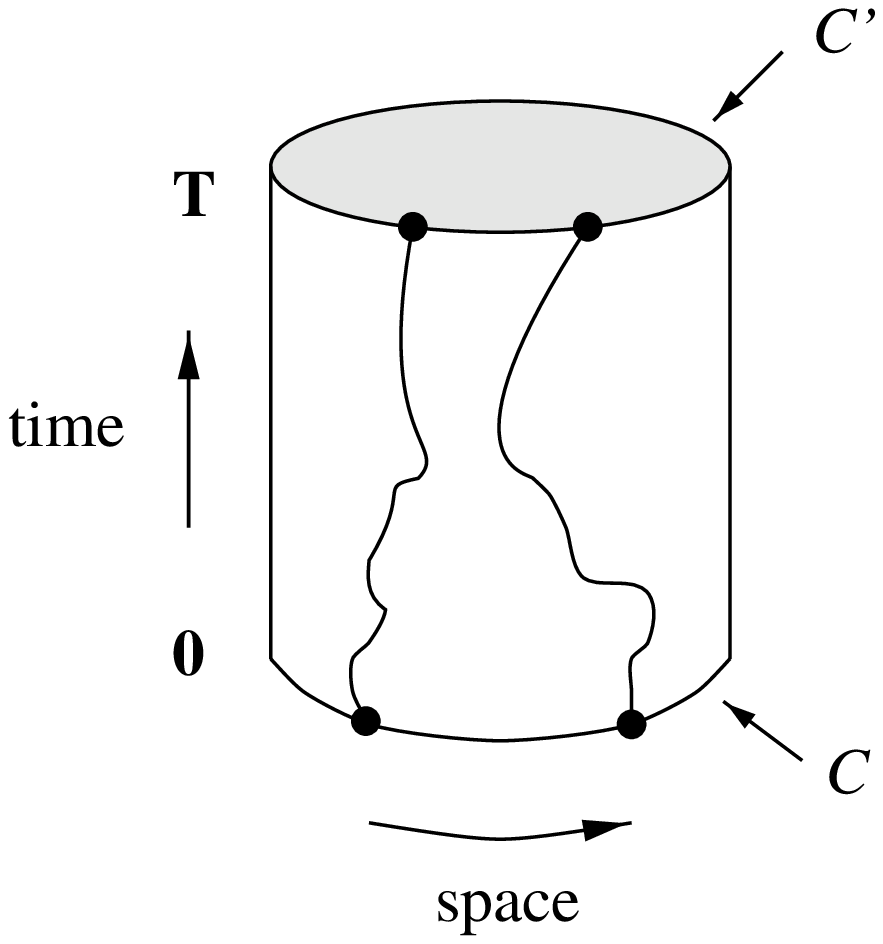,width=5.5cm,height=5.0cm}
\end{center}
\vspace{-0.0cm}
\caption{ \label{fig:WC}
\footnotesize Fermionic Wick contractions for the $f_{44}$ and $f_1$ correlators. 
$C$ and $C'$ denote the boundary gauge field respectively at $x_0=0$ 
and $x_0=T$; x denotes the insertion point of the local operator.}
\end{figure}

Our setup of lattice QCD is on a hyper-cubic 
euclidean lattice with spacing $a$ and size $L^3 \times T$.
We impose periodic boundary conditions in the spatial directions and Dirichlet
boundary conditions in time, as they are used to formulate the Schr\"odinger
functional (SF) \cite{sfscheme} 
(we refer to these references for unexplained notations).
Using homogeneous boundary conditions, 
where the spatial components
of the gauge potentials at the boundaries and also the fermion boundary fields
are set to zero, the Schr\"odinger functional partition 
function can be written as
\cite{ref:me_sf} 

\be
\cZ = \langle i_0 | \e^{-T\Ham} \Pgauge|i_0 \rangle\; ,
\ee
where the initial and final states $|i_0 \rangle$ carry the quantum 
numbers of the vacuum 
and $\Pgauge$ denotes a projector on gauge invariant states.
The states with charged pion quantum numbers 
in the Schr\"odinger functional are, indicating with $\zeta$ and $\bar\zeta$ 
(and the corresponding $\zeta'$ and $\bar\zeta'$)
a flavour doublet, the dimensionless fields
\be
\Source = \frac{a^6}{L^3} \sum_{\by,\bz} \bar\zeta(\by)\gamma_5 \tau^+ 
\zeta(\bz) \qquad \mbox{~and~}
\Source' = \frac{a^6}{L^3} \sum_{\bu,\bv} 
\bar\zeta'(\bu)\gamma_5 \tau^- \zeta'(\bv) \; ,
\label{eq:source_q}
\ee
where $\tau^{\pm}=\frac{1}{\sqrt{2}}(\tau^1 \pm i \tau^2)$ and 
$\tau^k$ with $k=1,2,3$ 
are the usual Pauli matrices.
The pion interpolating fields $\Source$ and $\Source'$ are respectively localized
at $x_0=0$ and $x_0=T$. 
The desired matrix element is obtained from the correlation function
\be
f_{44}(x_0) = -\frac{1}{2}\langle \;\Source' \cO_{44}(x) \;\Source \;\rangle\; ,
\ee
where we have used the independence on the spatial components $\bf{x}$ of $x=(x_0,\bf{x})$.
The Wick contractions of this correlation function contain also a disconnected piece that
we neglect consistently with the fact that we are interested on valence quark distribution.
For normalization purposes it is important to define
the boundary to boundary correlation function
\be
f_1 = -\frac{1}{2} \langle \;\Source' \;\Source \;\rangle\; .
\ee
The basic fermionic Wick contractions for these two correlation
functions are depicted in fig. \ref{fig:WC}.

In the following we will denote with $|i_0\rangle$ and $|i_\pi\rangle$
the states carrying respectively the quantum mumbers of the vacuum
and of the charged pion at zero momentum.
The correlation functions $f_{44}$ and $f_1$ have the following 
quantum mechanical
representations
\be
f_{44}(x_0) = 
\cZ^{-1} \frac{1}{2} \langle i_{\pi} | \e^{-(T-x_0)\Ham} \Pgauge \;\cO_{44}\; \e^{-x_0 \Ham} \Pgauge |i_{\pi} \rangle\; ,
\ee
\be
f_1 = \cZ^{-1}\frac{1}{2} \langle i_{\pi} |\e^{-T \Ham} \Pgauge |i_{\pi} \rangle
\; .
\ee
In order to extract the pion mass we have analyzed also the improved axial 
correlation function
\be
f_{A}^{I}(x_0) = -\frac{L^3}{2} \langle A_0^I(x) \; \Source \;\rangle\; ,
\ee
where
\be
A_0^I(x) = A_0(x) + ac_A\frac{1}{2}(\partial_0^* + \partial_0) P(x)\; ;
\ee
$c_A$ is the improvement coefficient
and the axial and pseudoscalar local operators take the form
\be
A_0(x) = \bar\psi(x) \gamma_0 \gamma_5 \tau^- \psi(x)
\ee
\be
P(x) = \bar\psi(x) \gamma_5 \tau^- \psi(x).
\ee
The non-perturbative value of $c_A$ was taken from ref. \cite{ref:csw_ca}.

Following ref. \cite{ref:me_sf} we insert a complete set of eigenstates 
of the hamiltonian and, retaining
only the first non-leading corrections, we have
for the improved axial current
correlation function  $f_A^{I}(x_0)$ and $f_1$,
\be
f_{A}^{I}(x_0) \simeq \frac{L^3}{2} \rho \langle 0,0| A_0^{I} |0,\pi \rangle 
e^{-m_{\pi }x_0} \{1 + \eta_A^{\pi} e^{-x_0\Delta} + \eta^{0}_A e^{-(T-x_0)m_G}\}
\label{fa}
\ee
\be
f_1 \simeq \frac{1}{2}\rho^2 \e^{-m_{\pi}T}\; .
\label{f1quantum}
\ee
where the coefficient appearing in this expansions are
\bea
   \rho &=&  {\langle 0,\pi | i_{\pi} \rangle
             \over 
             \langle 0,0 | i_0 \rangle} ,
                       \label{e_rho}\\
 \eta_A^{\pi} &=& {\langle 0,0| A^I_0 | 1,\pi\rangle \langle 1,\pi | i_{\pi} \rangle
                   \over 
                 \langle 0,0| A^I_0 | 0,\pi\rangle \langle 0,\pi | i_{\pi} \rangle}
                   \,, \\
 \eta^{0}_A &=& { \langle i_0 | 1,0 \rangle \langle 1,0| A_0  | 0,\pi\rangle 
                  \over
                 \langle i_0 | 0,0 \rangle \langle 0,0| A_0 | 0,\pi\rangle }
            \,.              
\eea
The energy gap in the pion channel between the fundamental 
and the first excited state is denoted by $\Delta$ 
and is estimated to be 
$\Delta r_0 \approx 3.2$, while $m_G$ is the mass of the $0^{++}$ glueball, 
$m_G r_0 \approx 4.3$ \cite{ref:me_sf}.
For the matrix element we find 
\be
f_{44}(x_0) \simeq \frac{1}{2} \rho^2 \langle 0,\pi| \cO_{44} | 0,\pi
\rangle e^{-m_{\pi }T}
\{1 + \eta_{\cO_{44}}^\pi e^{-x_0\Delta} + \eta_{\cO_{44}}^\pi
e^{-(T-x_0)\Delta}\}\; ,
\label{f44quantum}
\ee
where we define the ratio
\be
 \eta_{\cO_{44}}^{\pi} = {\langle 0,\pi| \cO_{44} | 1,\pi\rangle \langle 1,\pi | i_{\pi} \rangle
                   \over 
                 \langle 0,\pi| \cO_{44} | 0,\pi\rangle \langle 0,\pi | i_{\pi} \rangle}
                   \,.
\ee

\begin{figure}
\vspace{-0.0cm}
\begin{center}
\psfig{file=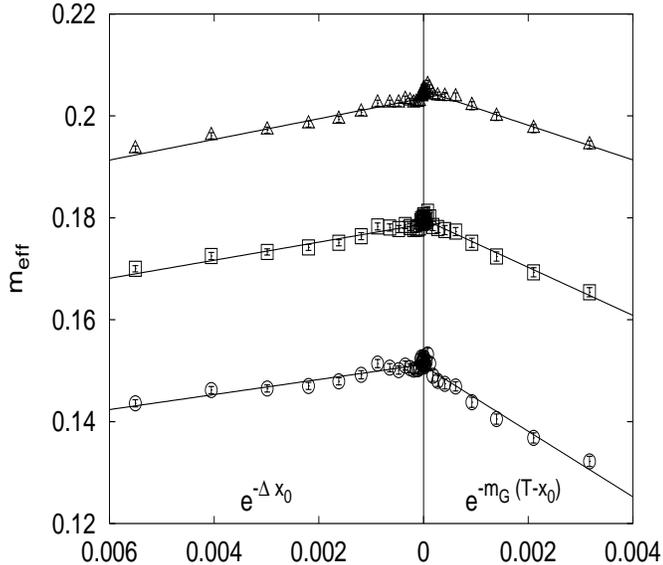,width=9cm,height=8.0cm}
\end{center}
\vspace{-0.0cm}
\caption{ \label{fig:effmass}
\footnotesize 
Effective pion mass as a function of the theoretically 
expected excited state 
contributions obtained from a transfer matrix decomposition.
Data corresponds to the simulation point at $\beta=6.45$ for three
$\kappa$ values with the non-perturbatively improved action 
(see table \ref{table:bare}).
In the left panel we plot
the effect of an excited pion state, whereas in the right panel 
we show the effects of the glueball contribution.
}
\end{figure}

A corresponding transfer matrix decomposition is obtained for the correlation
function $f_{12}$.
From these expressions it becomes clear that 
the matrix element we are interested in, neglecting contributions 
from excited states, 
can then be extracted from the plateau value of the following ratio:
\be
\frac{f_{44}(x_0)}{f_1} = \langle 0,\pi| \cO_{44} | 0,\pi \rangle 
     \; .
\ee
Finally, 
in order to relate this numerically computed ratio with the corresponding
continuum operators in Minkowski space, we need a suitable normalization
factor
\be
\langle x \rangle = \frac{2\kappa}{m_\pi} 
\langle 0,\pi| \cO_{44} | 0,\pi\rangle
\ee
with $\kappa$ the standard hopping parameter of the Wilson-Dirac-operator.
We remark here that $\langle x \rangle$ corresponds to the valence distribution
of a single quark ($u$ or $\bar d$ for example).
We followed ref.~\cite{ref:me_sf} in order to extract 
the plateau values for 
the effective pion mass and the matrix elements.  
Using the transfer matrix decomposition in eq.(\ref{fa}), 
the effects of higher excited states for the effective mass are given by
\be
m_{eff}(x_0) \simeq m_\pi { + \Delta \; \eta_A^\pi \;
e^{-\Delta \; x_0}}
{ - m_G \; \eta_A^0 \; e^{-m_G \; (T-x_0)}}\; .
\label{effmass}
\ee
In fig.~\ref{fig:effmass} we show the effective pion mass as function
of the anticipated excited state contaminations given in eq.~(\ref{effmass}). 
From the linear behaviour of the effective mass as a function of 
$e^{-\Delta \; x_0}$ and $e^{-m_G \; (T-x_0)}$  
we conclude that these are indeed the leading corrections.


\begin{figure}
\vspace{-0.0cm}
\begin{center}
\psfig{file=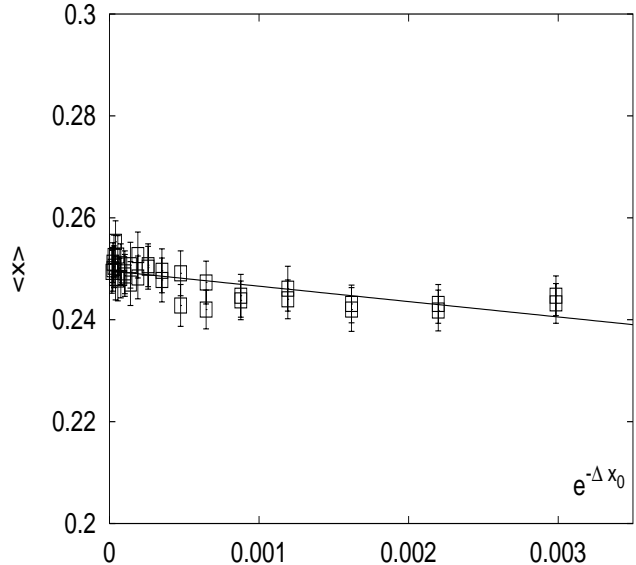,width=9cm,height=8.0cm}
\end{center}
\vspace{-0.0cm}
\caption{ \label{fig:effme}
\footnotesize
Matrix element as a function of the excited state 
contribution as obtained from a transfer matrix decomposition.
Data corresponds to the simulation point at $\beta=6.45$ for 
$\kappa = 0.1350$ with the non-perturbative improved action 
(see table \ref{table:bare}).
In the plot we superimpose the results for $x_0 > T/2$ and $x_0 < T/2$.
}
\end{figure}

The contribution from excited states  
on the plateau value of the matrix element following 
eqs. (\ref{f1quantum}) and (\ref{f44quantum}) is given by
\be
\langle x \rangle (x_0) \simeq \langle x \rangle
\left\{ 1 { + \eta_{\cO_{44}}^\pi \;
( e^{-\Delta \; x_0} + e^{-\Delta \; (T-x_0)})}\right\}\; .
\label{effplateau}
\ee
We show in fig.~\ref{fig:effme} the matrix element as a function 
of the expected excitation in eq.~(\ref{effplateau}). Again we observe
a linear behaviour of the matrix element indicating that  
the corrections come mainly from the first excited pion state. 
It is important to remark that $m_G$ and $\Delta$ have 
been computed using different boundary conditions (see ref. \cite{ref:me_sf}), 
where in principle excited states
corrections have different amplitudes. The agreement between 
the data and the expected form is then reassuring that the excited states
contamination is well controlled.


\begin{figure}
\vspace{-0.0cm}
\begin{center}
\psfig{file=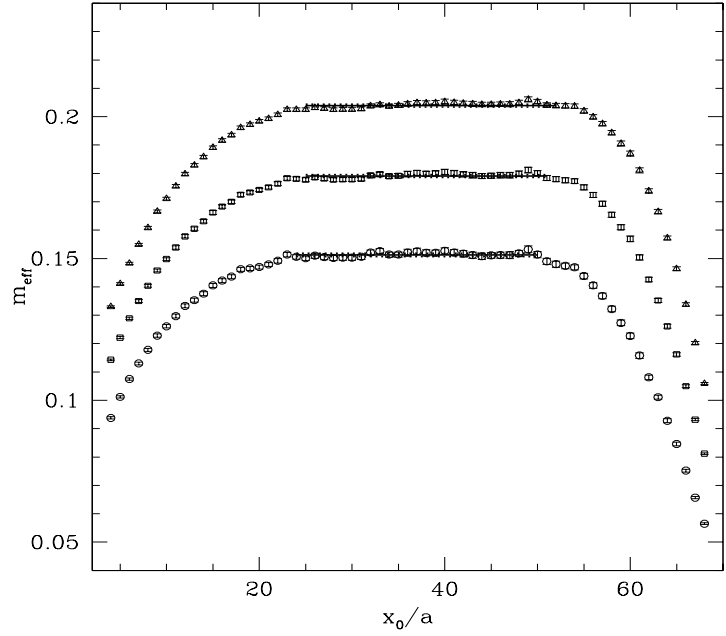,width=10cm,height=9.0cm}
\end{center}
\vspace{-0.0cm}
\caption{ \label{fig:mexample}
\footnotesize
Plateaux for the effective pion mass. The fit region to extract the mass
is indicated as a solid line. The simulation was done at 
$\beta=6.45$.
}
\end{figure}

\begin{figure}
\vspace{-0.0cm}
\begin{center}
\psfig{file=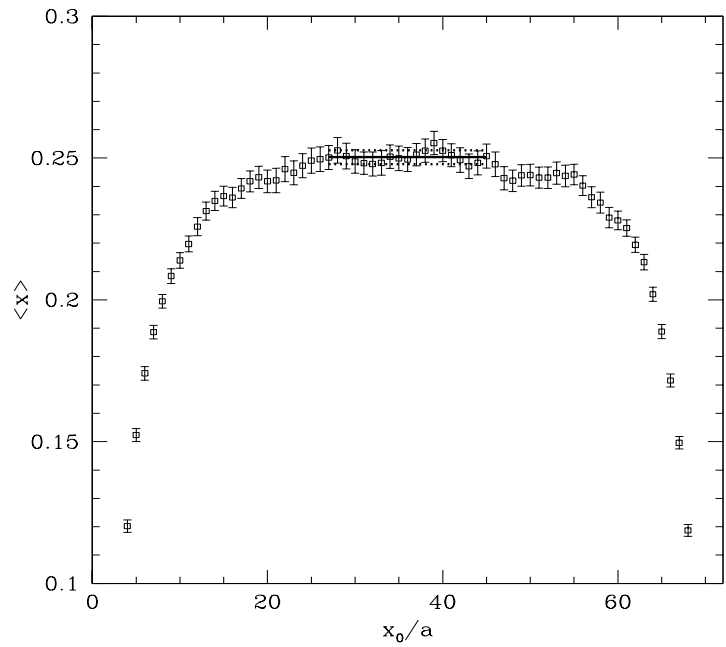,width=10cm,height=9.0cm}
\end{center}
\vspace{-0.0cm}
\caption{ \label{fig:maexample}
\footnotesize
Plateau for the matrix element. The fit region to extract the mass
is indicated as a solid line. The simulation was done at 
$\beta=6.45$.
}
\end{figure}

From fig.~\ref{fig:effmass}
and fig.~\ref{fig:effme}
we can read off, what is the systematic error on the pion mass or
the matrix element value if a particular value of the time separation
$x_0$ is taken to extract their values. 
In our analysis,
we demanded a 
systematic relative error, for all the $\beta$ and $k$ values,
coming from the excited states of 0.1\% for the pion mass
and  0.4\% for the matrix element which is well below 
the statistical
accuracy of our computations. 
Relating the value of $x_0$ to physical units, the above choice 
for our desired accuracy leads to a window for the extraction
of the plateau that for all the $\beta$ and $k$ values are around
\begin{itemize}
\item [ -- ] 1.2~fm $\lesssim x_0 \lesssim$ T-1.1~fm for $m_\mathrm{eff}$,
\item [ -- ] 1.3~fm $\lesssim x_0 \lesssim$ T-1.3~fm for the matrix element.
\end{itemize}
We give in fig.~\ref{fig:mexample} an example for the plateau behaviour
of the effective mass and in 
fig.~\ref{fig:maexample} an example for the plateau behaviour
of the matrix element. 
In table \ref{table:bare} we summarize the time intervals chosen for
$m_\pi$ and $\langle x \rangle$ that fulfill the aforementioned conditions.


\section{Numerical details and results}

In this section we will give numerical details and results about the computation
of the bare matrix element and of the renormalization constant at the low
energy scale $\mu_0$.

\subsection{Bare matrix element}

We have performed a set of quenched simulations for five $\beta$ values,
varying the lattice spacing between $a=0.093$ fm and $a=0.048$ fm, 
see table~\ref{table:parameters}.

\begin{table}[t]
\centering
\begin{tabular}{ccccl}
\hline
$\beta=6/g_0^2$ & $r_0/a$ & $L/a$ & $T/a$ & $~~N$ \\ \hline
6.0  &  5.37(2)  & 16 & 32 & ~600 \\
6.1  &  6.32(3)  & 24 & 42 & ~600 (Wilson)\\
6.1  &  6.32(3)  & 32 & 56 & ~185 (Clover)\\
6.2  &  7.36(3)  & 24 & 48 & ~600 \\
6.3  &  8.49(4)  & 24 & 64 & ~400 \\
6.45 & 10.46(5)  & 32 & 72 & ~400 \\ \hline
\end{tabular}
\caption{\footnotesize 
Parameters of our simulation points. $N$ denotes the number of measurements
taken into account. For all $\beta$-values except for $\beta=6.1$ the number
of measurements refers to Wilson and Clover simulations individually.}
\label{table:parameters}
\end{table}
In order to have a better control over the continuum extrapolation of
our lattice results we performed two independent sets of simulation at these
$\beta$ values, one employing standard Wilson fermions and the other using
non-perturbatively O(a)-improved Wilson fermions \cite{{ref:csw},{ref:csw_ca}}. 
As mentioned already 
above, we used the quenched approximation throughout this work. 
In order to achieve an extrapolation to the chiral limit, we employed 
three values of the quark mass corresponding to  
a value of $m_\pi$ that lies in the range
of 550 MeV - 1 GeV for all the $\beta$ values. 
For the simulations of the lightest quark mass at each $\beta$ value 
(the corresponding pion masses range in $4.5 \le m_\pi L \le 5.2$), 
we have corrected for the finite size effects, following ref. \cite{fse}.
For all the $\beta$ values the finite size corrections are in the region
$0.5\%-1.3\%$.
The summary of our bare results in table \ref{table:bare} are thus free from 
finite size effects.
The systematics coming from the higher excited states have been controlled
as explained in the previous section.


\begin{table}[h]
\centering
\begin{tabular}{ccccccc}
\hline
$\beta=6/g_0^2$ & $\kappa$ & $c_{sw}$ & Fit interval & $m_{\pi}$ & Fit interval & $\langle x \rangle$ \\\hline
6.0  & 0.153    & 0 & 14 - 21 & 0.4205(14) & 12 - 20 & 0.3080(18) \\
     & 0.154    & 0 & 14 - 20 & 0.3606(17) & 12 - 20 & 0.2994(24) \\
     & 0.155    & 0 & 14 - 19 & 0.2924(21) & 13 - 19 & 0.2876(38) \\\hline
6.1  & 0.151605 & 0 & 17 - 31 & 0.3685(05) & 17 - 25 & 0.3073(12) \\
     & 0.152500 & 0 & 17 - 30 & 0.3120(06) & 17 - 25 & 0.2951(16) \\
     & 0.153313 & 0 & 16 - 29 & 0.2536(07) & 16 - 26 & 0.2799(21) \\\hline
6.2  & 0.150600 & 0 & 19 - 35 & 0.3147(09) & 20 - 28 & 0.3037(21) \\
     & 0.151300 & 0 & 19 - 34 & 0.2674(10) & 20 - 28 & 0.2918(27) \\
     & 0.151963 & 0 & 18 - 33 & 0.2163(12) & 20 - 28 & 0.2763(38) \\\hline
6.3  & 0.149259 & 0 & 22 - 46 & 0.2968(08) & 22 - 42 & 0.3033(20) \\
     & 0.149978 & 0 & 22 - 45 & 0.2481(10) & 21 - 43 & 0.2888(29) \\
     & 0.150604 & 0 & 21 - 44 & 0.1996(12) & 19 - 45 & 0.2704(44) \\\hline
6.45 & 0.1486   & 0 & 29 - 51 & 0.2045(06) & 28 - 44 & 0.2854(23) \\
     & 0.1489   & 0 & 29 - 50 & 0.1808(07) & 28 - 44 & 0.2767(27) \\
     & 0.1492   & 0 & 30 - 49 & 0.1547(08) & 28 - 44 & 0.2652(35) \\\hline
6.0  & 0.1334 & NP \cite{ref:csw_ca}& 15 - 20 & 0.4021(13) & 14 - 18 & 0.2704(21) \\
     & 0.1338 & NP & 14 - 20 & 0.3551(13) & 14 - 18 & 0.2667(26) \\
     & 0.1342 & NP & 14 - 19 & 0.3028(16) & 15 - 17 & 0.2636(41) \\\hline
6.1  & 0.1340 & NP & 17 - 46 & 0.3534(05) & 16 - 40 & 0.2671(14) \\
     & 0.1345 & NP & 17 - 46 & 0.2947(05) & 16 - 40 & 0.2579(17) \\
     & 0.1350 & NP & 17 - 43 & 0.2239(06) & 16 - 40 & 0.2467(26) \\\hline
6.2  & 0.1346 & NP & 19 - 33 & 0.2798(07) & 19 - 29 & 0.2624(18) \\
     & 0.1349 & NP & 18 - 32 & 0.2430(08) & 19 - 29 & 0.2567(23) \\
     & 0.1352 & NP & 18 - 32 & 0.2008(09) & 18 - 30 & 0.2500(29) \\\hline
6.3  & 0.1346 & NP & 21 - 42 & 0.2640(07) & 22 - 42 & 0.2643(16) \\
     & 0.1349 & NP & 21 - 42 & 0.2284(07) & 22 - 42 & 0.2573(20) \\
     & 0.1352 & NP & 21 - 42 & 0.1881(08) & 21 - 43 & 0.2467(26) \\\hline
6.45 & 0.1348 & NP & 26 - 52 & 0.2040(05) & 26 - 46 & 0.2566(19) \\
     & 0.1350 & NP & 25 - 51 & 0.1791(05) & 26 - 45 & 0.2502(24) \\
     & 0.1352 & NP & 25 - 50 & 0.1513(06) & 27 - 45 & 0.2426(31) \\\hline
\end{tabular}
\caption{\footnotesize 
Results for the pseudoscalar mass and the bare matrix element at 
all our simulation points. We specify also the fit interval for the 
effective mass and the bare matrix element obtained from the request
of having the systematic uncertainty coming from the excited states
below $0.1\%$ and $0.4\%$, respectively. See text for further details.}
\label{table:bare}
\end{table}

\subsection{Renormalization constant}
\label{sec.:3.2}

In order to renormalize the bare matrix element at the scale $\mu_0$, where
we can make contact to the  
running described by the non-perturbatively
computed UV invariant 
step scaling function
$\EuFrak{S}_{\rm INV}^{\rm UV,SF}(\mu_0)$,
we have to compute the renormalization
constants $Z^{\rm SF}(\mu_0)$. 
The continuum limit of the renormalized matrix element 
requires 
to compute $Z^{\rm SF}(\mu_0)$
at exactly the lattice spacing, where the matrix element has been 
calculated, while keeping the scale fixed. Decreasing the lattice spacing 
$a$, we would 
hence have to 
increase the lattice volume
in order to stay at a fixed value of $\mu_0=(1.436 r_0)^{-1}$. Since we 
can not vary the lattice size continuously, we 
performed instead simulations on a sequence of lattice sizes and adjusted 
the values of $\beta$ to realize the correct value of $\mu_0$.
The values of $\beta$ are slightly different from the ones used in ref. \cite{alfa_mass},
and were obtained \cite{ref:rainer_priv} using a new determination of 
$r_0/a$ \cite{ref:silvia_rainer}.
We recall that for our determination of the renormalization constant at the
scale $\mu_0$ using the finite volume SF scheme we have performed 
simulations directly at $\kappa=\kappa_c$, determined through the PCAC Ward identity.
Details about the computation of the renormalization factor, such as 
renormalization condition and external parameters typical of the SF scheme,
can be found in ref. \cite{rgissf}.
We give in table~\ref{table:Zmu0} the parameters of our runs and the values 
of the
$Z$ factors at $\mu_0$ for both Wilson and O(a)-improved Wilson fermions.

\begin{table}[t]
\centering
\begin{tabular}{cccccc}
\hline
$L/a$ & $\beta=6/g_0^2$ & $\kappa_c$ & $c_{sw}$ & $Z_{\cO_{44}}^{SF}$ & $Z_{\cO_{12}}^{SF}$\\\hline
 8    &      6.0055     &  0.153597(14)  &  0   &  0.3211(44)   &  0.3673(35)    \\ 
10    &      6.1425     &  0.152277(10)  &  0   &  0.2994(43)   &  0.3467(35)    \\
12    &      6.2670     &  0.151024(7)   &  0   &  0.3008(43)   &  0.3462(33)    \\
16    &      6.4825     &  0.149012(12)  &  0   &  0.2861(57)   &  0.3262(44)    \\ \hline
 8    &      6.0055     &  0.135006(5)   &  NP  \cite{ref:csw_ca}&  0.3451(37)   &  0.3423(31)    \\
10    &      6.1425     &  0.135625(4)   &  NP  &  0.3204(36)   &  0.3260(31)    \\
12    &      6.2670     &  0.135756(3)   &  NP  &  0.3131(35)   &  0.3287(30)    \\
16    &      6.4825     &  0.135617(4)   &  NP  &  0.3029(43)   &  0.3167(37)    \\ \hline
\end{tabular}
\caption{\footnotesize 
Results for $Z_{\cO_{44}}^{SF}$ and $Z_{\cO_{12}}^{SF}$ at the scale $\mu_0 = (1.436r_0)^{-1}$
with Wilson and non-perturbatively improved clover actions.}
\label{table:Zmu0}
\end{table}

\begin{figure}
\vspace{-0.0cm}
\begin{center}
\psfig{file=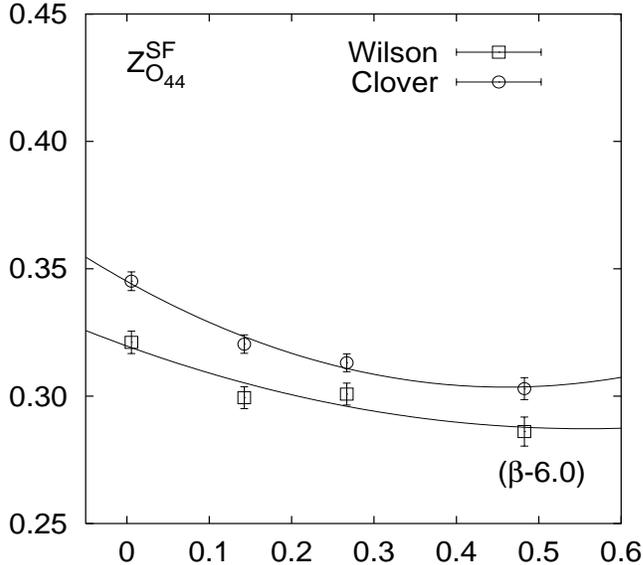,width=12cm,height=8.0cm}
\end{center}
\vspace{-0.0cm}
\caption{ \label{fig:zint}
\footnotesize
Numerical results for $Z^{\rm SF}_{\cO_{44}}(g_0,\mu_0)$ 
together with their interpolating polynomials.}
\end{figure}

In order to match the $\beta$ values where the matrix elements themselves 
have been computed, we describe the $\beta$ dependence by an effective
interpolation formula,
\be
Z^{{\rm{SF}}}(\mu_0) = \sum_{i=0}^2 z_i^{{\rm{SF}}}(\beta - 6.0)^i 
\label{zmu0}
\ee
with the coefficients for Wilson and 
O(a)-improved Wilson fermions listed in table \ref{table:interp}.
The statistical uncertainty to be taken into account when using this formula
is about $1.1\%$ for the non-perturbatively improved clover action and 
$1.4\%$ for the Wilson action \footnote{For $\beta \approx 6.5$ the error to be 
associated grows to $1.4\%$ for the non-perturbatively improved clover action
and to $2.0\%$ for the Wilson action.}.
In fig. \ref{fig:zint} we show the $\beta$ dependence of $Z^{{\rm{SF}}}(\mu_0)$ 
for the $\cO_{44}$ operator and for the two actions used, together
with the plot of the interpolating formula (\ref{zmu0}).

\begin{table}[t]
\centering
\begin{tabular}{ccccc}
\hline
$c_{sw}$ & applicability & i & $z_i^{SF}$ & $z_i^{RGI}$ \\\hline
 0                &  $6.0\le \beta \le 6.5$ & 0 &  \phantom{-}0.3197  &  \phantom{-}1.446 \\
                  &                         & 1 &            -0.1166  &            -0.527 \\
                  &                         & 2 &  \phantom{-}0.1046  &  \phantom{-}0.473 \\\hline
 NP \cite{ref:csw_ca} &  $6.0\le \beta \le 6.5$ & 0 &  \phantom{-}0.3451  &  \phantom{-}1.561 \\
                  &                         & 1 &            -0.1806  &            -0.817 \\
                  &                         & 2 &  \phantom{-}0.1962  &  \phantom{-}0.888 \\\hline

\end{tabular}
\caption{\footnotesize 
Coefficients of the interpolating polynomials, eq. (\ref{zmu0}) and (\ref{zrgi}).
Uncertainties are discussed in the text.}
\label{table:interp}
\end{table}

Using this interpolation formula, we can match the values of $a$
corresponding to the $\beta$ values used for the computation
of the bare matrix element,
which allows finally to obtain 
the renormalized matrix element at the 
scale $\mu_0$ in the continuum limit. 

\begin{figure}[t]
\begin{center}
\hskip-4mm
\epsfig{file=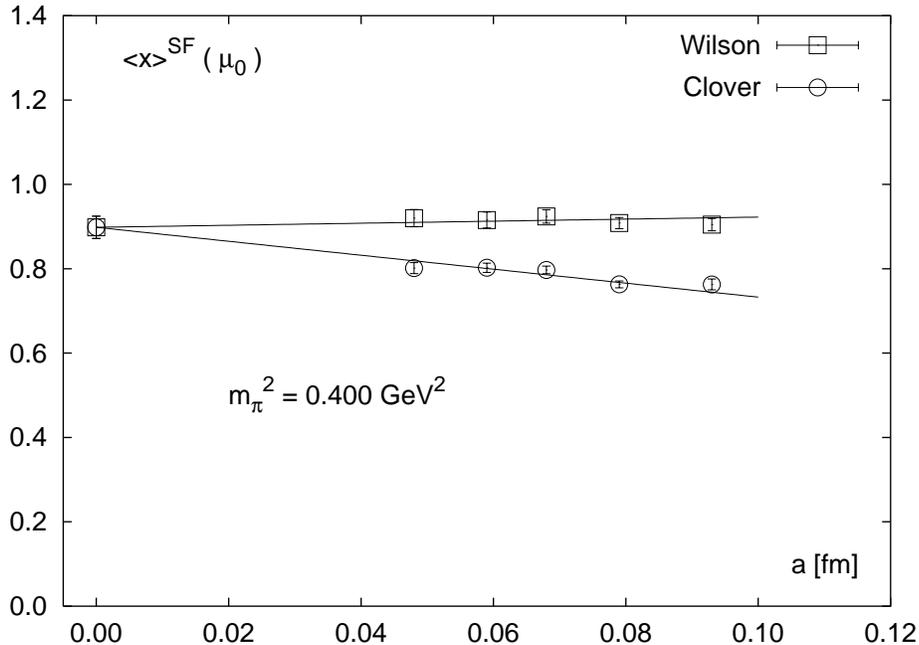,height=9.0cm}
\vskip-6mm
\end{center}
\caption{
\footnotesize
Combined continuum extrapolation for
the pion matrix element computed with Wilson and O(a)-improved 
Wilson fermions, at $(r_0m_\pi)^2 = 2.57$, which corresponds to a pion
mass of $m_\pi = 632$ MeV.}
\label{fig:cont_X}
\end{figure}

We have performed a continuum extrapolation at fixed values of
$(r_0m_\pi)^2$. Fixing $(r_0m_\pi)^2$ is achieved by interpolating our bare matrix elements 
linearly in the quark mass.
The physical values of the pion mass range $550 $ MeV $ < m_\pi < 1$ GeV.
In fig.~\ref{fig:cont_X} we show an example of the continuum limit 
at our next to lowest pion mass ($(r_0m_\pi)^2 =2.57$). The continuum extrapolation
of the renormalized matrix element

\be
\langle x \rangle^{SF} (\mu_0,r_0m_\pi) = \lim_{a\to 0}
\frac{\langle\pi|{\cal O}_{44}|\pi\rangle(a,r_0m_\pi)}{Z^{{\rm{SF}}}(a,\mu_0)} \; ,
\ee
has been done via a constrained linear fit in the lattice spacing $a$ 
using simulation results obtained with the Wilson and clover action.
For all the values of the pion mass we have performed this constrained
extrapolation excluding the coarsest lattice of our data set.
In section \ref{sec:che} we will discuss the uncertainties connected with 
the continuum extrapolation, and the one related to the chiral extrapolation.
If we perform the chiral extrapolation linearly in the
pion mass squared directly in the continuum 
we obtain a value of
\be
\langle x \rangle^{SF} (\mu_0) = 0.810(33)
\label{mesf}
\ee
for the continuum renormalized pion matrix element in the SF scheme.

We can anticipate here for clarity that, in this particular case,
to invert the order of the chiral and continuum extrapolation 
gives a fully consistent result, independently on how the continuum
extrapolation is performed.

\section{RGI matrix element}

The phenomenological analysis of the experimental data
is usually done in the $\msb$ scheme. In order to translate our fully
non-perturbatively computed matrix element
$\langle x \rangle^{SF}(\mu_0)$ renormalized in the SF scheme to
the $\msb$ scheme,
we need first to calculate the universal renormalization group invariant
matrix element. This is done following the complete non-perturbative
evolution \cite{rgissf} of the step scaling function in the SF scheme 
as outlined above.
Using the UV invariant step scaling function as determined in 
our earlier work \cite{rgissf}, it is now possible
to eliminate any reference to the scale $\mu_0$ and to obtain finally the
RGI matrix element. In ref. \cite{rgissf} we have already computed

\be
\EuFrak{S}^{UV,SF}_{INV}(\mu_0) = 0.221(9)\; .
\label{eq:siguvinv}
\ee

Thus using formula (\ref{pieces_of_oren}) we can obtain the main result of this paper
\be
{\langle x \rangle}_{RGI} = 
{\langle x \rangle}^{SF}(\mu_0)\;\EuFrak{S}^{UV,SF}_{INV}(\mu_0) = 0.179(11) \; .
\ee

The RGI matrix element allows a simple conversion to any desired scheme
(e.g.$\;\msb$ at 2 GeV) requiring only the knowledge of the $\beta$- and
$\gamma$-function up to a certain order in perturbation theory in this scheme
\cite{{b2_msbar},{g1_msbar}}.
The total renormalization factor to directly translate any bare matrix element
of $\cO_{44}$ into the RGI matrix element, can be written 
as 
\be
Z_{\rm RGI}(g_0) = \frac{Z^{\rm SF}(g_0,\mu_0)}{\EuFrak{S}^{UV,SF}_{INV}(\mu_0)}\; .
\ee
Combining eq. (\ref{eq:siguvinv}) and the interpolating formula (\ref{zmu0})
we obtain a further interpolating polynomial  
\be
Z_{{\rm{RGI}}} = \sum_{i=0}^2 z_i^{{\rm{RGI}}}(\beta - 6.0)^i 
\label{zrgi}
\ee
whose coefficients are listed in table \ref{table:interp}. 
These parametrizations of $Z_{\rm RGI}$ are to be used with the same 
uncertainty of $Z^{{\rm{SF}}}$ and an additional error of $4.0\%$,
coming from the uncertainty of $\EuFrak{S}^{UV,SF}_{INV}(\mu_0)$,
that has to be added quadratically after performing a continuum
extrapolation.
Our resulting number $\langle x \rangle^{\msb} (\mu$=2GeV)=0.246(15)
is not fully compatible with previous lattice
computation \cite{ref:best}, where it was found that 
$\langle x \rangle^{\msb}(\mu$=2.4GeV)=0.273(12), and in better agreement with
the result coming from the global fits \cite{experiment} of experimental data 
$\langle x\rangle_{phen}(\mu=2\mathrm{GeV}) = 0.21(2)$.
The disagreement of these numbers could be explained by the 
fact that in this paper we analyze and correct for several sources of 
systematic errors. We apply a non-perturbative renormalization 
and perform the continuum limit, while in \cite{ref:best} a perturbative 
renormalization was adopted without performing the continuum limit 
(only one lattice spacing).
Moreover we correct, where present, for finite size effects.

\section{Chiral and continuum extrapolation}
\label{sec:che}

\begin{figure}
\vspace{-0.0cm}
\begin{center}
\psfig{file=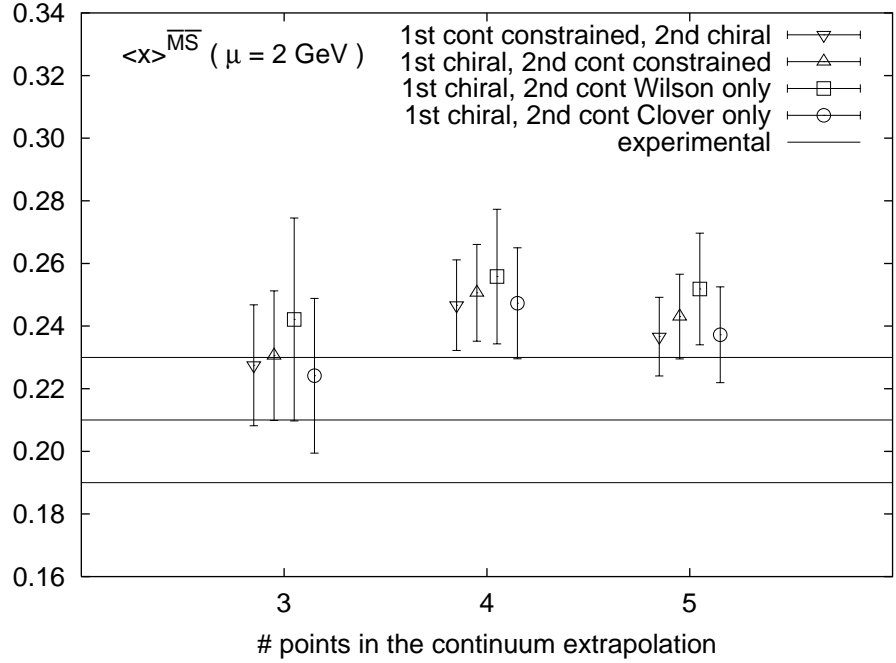,width=9cm,angle=270}
\end{center}
\vspace{-0.0cm}
\caption{ \label{fig:cont_exp}
\footnotesize
Value of $\langle x \rangle^{\msb} (\mu=2$ GeV), obtained
from our lattice simulations, as
a function of the number of points used to perform the continuum
extrapolation.
For each choice of the number of points used we compare several
ways to perform the continuum and chiral limit: 
({\large$\triangledown$}) first the constrained continuum limit and then the chiral extrapolation; 
($\triangle$) first the chiral extrapolation and then the constrained continuum limit; 
($\square$) first the chiral extrapolation and then the continuum limit using only
the Wilson data; 
({\Large$\circ$}) first the chiral extrapolation and then the continuum limit using only
the clover data.
The band represents the same quantity, with its error, obtained through 
global fits of the experimental data \cite{experiment}.
}
\end{figure}

Since the form of the correct chiral extrapolation of the lattice data
obtained at rather large quark masses is still under debate
(see \cite{panel2001} for a recent review and refs. therein, while for
the problem considered here see \cite{{thomas1},{australians},{savage1},{savage2},{savage3},{chen1}} ), we aimed at avoiding a possible 
source of systematic error and 
{\em first} performed a continuum extrapolation and {\em then} tried to 
extrapolate the data to the chiral limit. 
In this way, continuum chiral perturbation theory
is applicable and we do not have to worry that chiral symmetry breaking
effects may spoil the chiral extrapolation at non-vanishing values of
the lattice spacing. 
We fixed the physical values of the 
pion mass in units of $r_0$ \cite{ref:r0} and performed a continuum 
extrapolation keeping $r_0 m_\pi$ fixed. To this end
we had to slightly interpolate the values of the matrix elements, since
in our simulations the pion mass at different values of $\beta$ were not
obtained at the same value of the physical pion mass.  
The corresponding error, taking the correlations into
account, of this slight interpolation is, however, well below the 
statistical accuracy of our numerical data.
We have performed the continuum limit of the matrix element from our 
simulation points using three, four and five values of $\beta$,
employing Wilson and O(a)-improved Wilson fermions, in order
to study the effect  of including
coarser lattices for the continuum extrapolation.
This detailed analysis is summarized in fig. \ref{fig:cont_exp}.
In all the following discussion, when we talk about chiral extrapolation,
we use a linear extrapolation in the squared pion mass.


In this figure we plot the value of $\langle x \rangle^{\msb} (\mu=2$ GeV) as
a function of the number of points used to perform the continuum
extrapolation.
Moreover, for each choice of the number of points used, we compare several
ways to perform the continuum and chiral limit (the symbols refer to the 
symbols in fig. \ref{fig:cont_exp}): \\
1) first the constrained continuum limit and then the chiral extrapolation
({\large$\triangledown$}); \\
2) first the chiral extrapolation and then the constrained continuum limit ($\triangle$); \\
3) first the chiral extrapolation, then the cont.~limit using only
the Wilson data ($\square$); \\
4) first the chiral extrapolation, then the cont.~limit using only
the clover data ({\Large$\circ$}).

All these lattice results are compared with the same quantity obtained through 
global fits of the experimental data \cite{experiment}.
Our conclusions, for this quantity and for the range of masses and lattice
spacings simulated are twofold. First of all, the order of how to perform the
chiral and continuum extrapolation is irrelevant, independently from which action is
used to do the continuum limit.
Then the continuum limit is consistent if performed using three, four or five
points, also in this case, independently of the lattice action used. 

We decided to perform first the continuum limit at fixed pion mass using the
four points corresponding to the four smallest lattice spacings. 
Only as a second step we then perform the chiral extrapolation linearly
in the squared pion mass. The continuum extrapolated data at fixed $m_\pi^2$
are collected in table \ref{table:continuum}.


\begin{table}[t]
\centering
\begin{tabular}{cc}
\hline
$m_\pi^2$[GeV$^2$] & $\langle x \rangle^{\msb}$ ($\mu = 2$ GeV)\\\hline
  1.000   & 0.312(20) \\
  0.900   & 0.306(17) \\
  0.800   & 0.300(15) \\
  0.700   & 0.293(14) \\
  0.600   & 0.285(13) \\
  0.500   & 0.279(13) \\
  0.400   & 0.274(14) \\
  0.315   & 0.269(15) \\ \hline
  0.000   & 0.246(15) \\ \hline
\end{tabular}
\caption{\footnotesize 
Results for the matrix element in the continuum limit (quenched approximation),
converted to the $\msb$ scheme at 2 GeV. The continuum extrapolation was
performed with a constrained linear fit of the Wilson and non-perturbatively
improved clover data for the four smaller lattice spacings. The value in the
chiral limit is obtained from a linear fit (see text for details).}
\label{table:continuum}
\end{table}

A remark is in order here. It has been shown in ref. \cite{savage1}
that in quenched chiral perturbation theory (Q$\chi$PT) 
at the leading order, this particular matrix element is free
from chiral logarithms. In ref. \cite{savage3} it has been shown that in the full
$\chi$PT indeed there are chiral logarithms. 
In ref. \cite{australians} an effective phenomenogical ansatz is given by
\be
\langle x^n\rangle = A_n \left[ 1-\frac{1}{(4\pi f_\pi)^2}
                     m_\pi^2\ln
                     \left(\frac{m_\pi^2}{m_\pi^2 + \Lambda^2 }
                     \right)\right]
                     +B_n m_\pi^2\; ,
\label{chiralfit}
\ee
motivated by the chiral expansion in the full theory. A similar formula was
suggested to explain the discrepancy between the lattice results and the global
fits of the experimental data of the proton $\langle x \rangle$.


\begin{figure}
\vspace{-0.0cm}
\begin{center}
\psfig{file=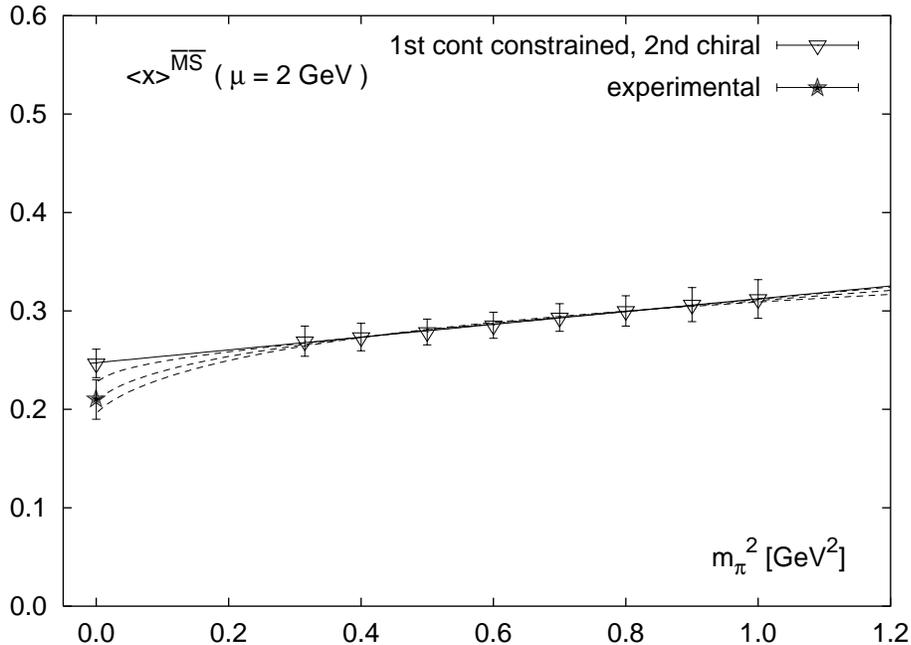,height=9.0cm}
\end{center}
\vspace{-0.0cm}
\caption{ \label{fig:chiralcont}
\footnotesize
Chiral extrapolation {\em in the continuum}. The dashed curves refer to
the formula (\ref{chiralfit}) with $\Lambda=\{0.4,0.7,1.0\}$ GeV and $f_\pi=93$ MeV.
The chiral extrapolated value ({\large$\triangledown$}) is compared 
with the phenomenological estimate ({\large$\star$}).
}
\end{figure}

In fig.~\ref{fig:chiralcont} we show the {\em continuum} values
of the renormalized matrix elements as a function of the pion mass
(see table \ref{table:continuum}). In order to
illustrate possible unquenching effects we also perform a chiral
extrapolation using eq.(\ref{chiralfit}). For the different dashed 
curves in fig.~\ref{fig:chiralcont} we used different values of the 
parameter $\Lambda=\{0.4,0.7,1.0\}$ GeV that is supposed to provide an estimate of the size
of the pion cloud. However, the data, even in the continuum limit, clearly
prefer a straight line extrapolation consistent with the results of Q$\chi$PT. 
Using a linear chiral extrapolation we obtain a final value for our matrix
element consistent with the experimental results, up to quenching effects.
In order to disentangle the quenched and chiral uncertainties,
lattice data have to be provided very close to the physical point,
where the pion assumes a mass as measured in experiment; presumably
chiral invariant formulations of QCD are necessary to clarify the 
question of the chiral extrapolation. For the time being and for this 
work we take the straight line behaviour as suggested by the numerically
obtained data for the chiral extrapolation. 

\subsection{Discussion and concluding remarks}
\label{sec.:5.1}

The main results of our paper, i.e. the renormalization 
group invariant matrix element and the renormalized matrix 
element at a scale of 2 GeV in the $\msb$ scheme,
are summarized in eqs.~(\ref{rgiresult},\ref{finresults}). 
Comparing our results to phenomenological estimates, 
it comes out to be slightly higher, but already more precise than the latest NLO
analyses of Drell-Yan and prompt photon $\pi N$ data \cite{experiment},
which yield a combined experimental value of
$\langle x \rangle^{\msb} (\mu$=2GeV) = 0.21(2).
These results have been computed for the valence quark distribution
\footnote{We remark here that,
as stated in ref. \cite{experiment}, the extraction
of the sea distribution in a pion is currently impossible 
due to the lack of the necessary data at low Bjorken $x$.}.
We conclude that the non-perturbatively obtained number
for the twist-2 matrix element of a pion is not completely
consistent with phenomenological estimates. 
Although this should not be too worrisome, since we are still left 
with the quenched approximation, 
it is worthwhile to discuss the lattice numbers further. 

One cause for the deviation could lie in the chiral extrapolation 
of this matrix element that
was done linearly in the pion mass. If we would use the
phenomenological fit ansatz of ref.~\cite{australians}, we 
find $\langle x\rangle^{\msb}(2\mathrm{GeV})= 0.21(1)(2)$, where the first error corresponds to
varying the data in their errorbars and the second error
is the variation for $\Lambda$ in the range 0.4 to 1.0 GeV.
Clearly this chiral extrapolation would 
reconcile the lattice results with experiments. However, as can be seen
in fig.~\ref{fig:chiralcont}, at the moment it is not possible to test such an
ansatz unambiguously, since the values of quark masses used in the
simulations are too large. Only with future simulations, presumably 
employing e.g.
chiral invariant formulations of lattice QCD, this question could be
clarified. 

Another cause of the deviation could certainly be 
the quenched approximation, which is a severe limitation and effects
of dynamical quarks have to be explored
in the future.
In any case, we conclude that 
the Schr\"odinger functional method has been proven to provide an excellent 
tool to compute the non-perturbative scale evolution of 
multiplicatively renormalized quantities such as the quark mass 
\cite{alfa_mass} or the matrix elements \cite{rgissf} considered
in this work. 

In order to further utilize our results, we attempted to 
use the non-perturbatively obtained renormalization constants to see
the effect of a non-perturbative renormalization on the results of
other groups on the quenched bare matrix element of the $\cO_{44}$ operator between 
proton states (preliminary results can be found in \cite{ref:andrea_tsukuba}).

\begin{figure}
\vspace{-0.0cm}
\begin{center}
\psfig{file=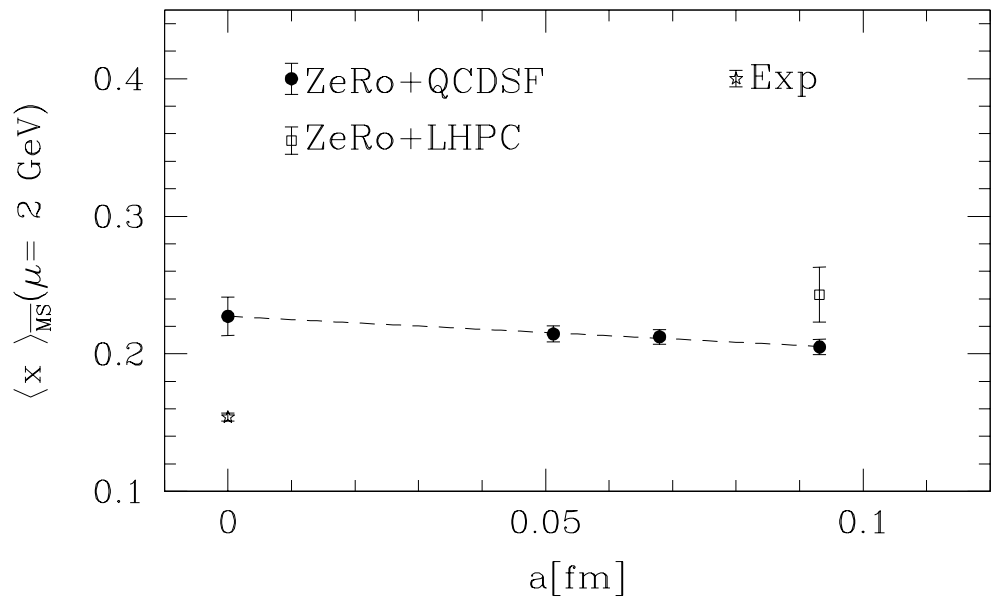,width=15cm,height=15.0cm}
\end{center}
\vspace{-6.5cm}
\caption{ \label{fig:x_prot}
\footnotesize
Continuum extrapolation of the non-perturbatively renormalized matrix element 
of $\cO_{44}$ between proton states based on the quenched bare data, 
both already chirally 
extrapolated linearly in the quark mass, 
from refs. \cite{QCDSF_priv} and \cite{LHPC}, using the non-perturbatively
computed Z factor of the present work (ZeRo coll.)}
\end{figure}

The LHPC coll. \cite{LHPC} has computed the bare matrix element 
of the $\cO_{44}$ operator between proton states at $\beta=6.0$ using
the Wilson action.
The QCDSF coll. \cite{QCDSF_priv} has computed the bare matrix element 
of the $\cO_{44}$ operator between proton states at $\beta=6.0,6.2,6.4$ using
the non-perturbatively improved clover action.
Both the collaborations perform a chiral extrapolation linear in the
quark mass at fixed lattice spacing.
If we take now the non-perturbative renormalization factor computed in this
paper (ZeRo coll.), we obtain the result summarized in fig. \ref{fig:x_prot},
where we show the continuum extrapolation 
of the non-perturbatively renormalized matrix element 
of $\cO_{44}$ between proton states based on the bare data, already chirally 
extrapolated, from refs. \cite{QCDSF_priv}, and as a comparison the same 
non-perturbatively renormalized matrix element using the bare data of ref. 
\cite{LHPC}.
In fig. \ref{fig:x_prot} we also indicate the result obtained from global fits
of the experimental data as listed in ref. \cite{LHPC} (see refs. inside for
details).
The final numerical results are
\bea
\langle x \rangle^{\msb}(\mu=2 {\rm GeV}) &=& 0.227(14) ~~  {\rm ZeRo+QCDSF} \\
\langle x \rangle^{\msb}(\mu=2 {\rm GeV}) &=& 0.243(20) ~~  {\rm ZeRo+LHPC}~~
\beta=6.0 \\
\langle x \rangle^{\msb}(\mu=2 {\rm GeV}) &=& 0.154(3) ~~~~  {\rm global~fits}
\eea
For the first time a continuum extrapolation 
in the case of the proton matrix element could be done, thanks
to the non-perturbative renormalization factor computed in this paper.

\subsection*{Acknowledgments}
We thank S. Capitani for many useful discussions.
We thank R. Sommer for sending us the updated values of $\beta$ 
at the matching scale used in this work, and we thank G. Schierholz
for sending us, prior to publication, the values of the proton
bare matrix element.       
The computer center at NIC/DESY Zeuthen provided the necessary technical 
help and the computer resources. 
This work was supported by
the EU IHP Network on Hadron Phenomenology from Lattice QCD
and by the DFG Sonderforschungsbereich/Transregio SFB/TR9-03.

\begin{appendix}

\section*{Appendix}
For the computation of the matrix elements the quark fields are chosen
to be periodic in the three space directions,
\be
  \psi(x+L\hk)=\psi(x),
  \qquad
  \psibar(x+L\hk)=\psibar(x)\; .
\label{eq:tetadef}
\ee
The lattice derivatives in the forward direction are given by

\be
  \nabla_\mu\psi(x)=\frac{1}{a}[
  U(x,\mu)\psi(x+a\hmu)-\psi(x)]
\label{eq:nab_right}
\ee
\be
  \nabla_\mu^*\psi(x)=\frac{1}{a}[
  \psi(x) - U(x-a\hmu,\mu)^{-1}\psi(x-a\hmu)]
\label{eq:nabstar_right}
\ee

and the backward derivatives are defined by

\be
  \psibar(x) \overleftarrow \nabla_\mu = 
  \frac{1}{a}[\psibar(x + a \hmu)U(x,\mu)^{-1} - \psibar(x)]
\ee

\be
   \psibar(x) \overleftarrow \nabla_\mu^{*} = 
   \frac{1}{a}[\psibar(x) - \psibar(x - a \hmu)U(x - a\hmu,\mu)]\; .
\ee

\end{appendix}

\input refs

\end{document}

%% file: refs
\newpage

\def\NPB #1 #2 #3 {Nucl.~Phys.~{\bf#1} (#2)\ #3}
\def\NPBproc #1 #2 #3 {Nucl.~Phys.~B (Proc. Suppl.) {\bf#1} (#2)\ #3}
\def\PRD #1 #2 #3 {Phys.~Rev.~{\bf#1} (#2)\ #3}
\def\PLB #1 #2 #3 {Phys.~Lett.~{\bf#1} (#2)\ #3}
\def\PRL #1 #2 #3 {Phys.~Rev.~Lett.~{\bf#1} (#2)\ #3}
\def\PR  #1 #2 #3 {Phys.~Rep.~{\bf#1} (#2)\ #3}

\def\etal{{\it et al.}}
\def\ibid{{\it ibid}.}